# Structures of Spherical Viral Capsids as Quasicrystalline Tilings

O.V. Konevtsova,[*] V.L. Lorman,[†] and S.B. Rochal[*]

[*]Faculty of Physics, Southern Federal University, 5 Zorge str., 344090 Rostov-on-Don, Russia
[†]Laboratoire Charles Coulomb, UMR 5221 CNRS and Universite´ Montpellier 2, pl. E. Bataillon, 34095 Montpellier, France

**Abstract**—Spherical viral shells with icosahedral symmetry have been considered as quasicrystalline tilings. Similarly to known Caspar–Klug quasi-equivalence theory, the presented approach also minimizes the number of conformations necessary for the protein molecule bonding with its neighbors in the shell, but is based on different geometrical principles. It is assumed that protein molecule centers are located at vertices of tiles with identical edges, and the number of different tile types is minimal. Idealized coordinates of nonequivalent by symmetry protein positions in six various capsid types are obtained. The approach describes in a uniform way both the structures satisfying the well-known Caspar–Klug geometrical model and the structures contradicting this model.

## 1. INTRODUCTION

It is well known that host cell infection by a virus strongly depends on the protein arrangement in the capsid [1–3], i.e., the solid viral shell consisting of molecules of the same type (more rarely, of several types) and protecting the virus genome from external influences. This arrangement is regular, symmetric, and has a high degree of positional and orientational ordering of protein molecules. However, despite certain similarities in the capsid and classical crystal structures, the generalization of the concepts of solid state physics, giving insight into the virus structure was initiated rather recently. An analysis of the capsid structures goes back to the pioneering study by Crick and Watson in 1956 [4], who asserted that all small viruses are constructed from a limited number of identical protein molecules packed together on a regular basis with cubic symmetry. In 1960, Caspar and Klug (CK) obtained the first experimental data indicating icosahedral symmetry of the viral capsid [5]; in 1962, the same authors proposed a geometrical model based on the interpretation of the viral capsid structure in terms of the icosahedron net whose faces are filled with periodic hexagonal motif [6]. Assembly of a three-dimensional structure from a plane one results in the appearance of twelve convex topological defects located at icosahedral vertices, i.e., pentamers corresponding to cut_out 60° sectors of the icosahedron net. Net edges are connected continuously, since cut-out sector edges before joining were equivalent and were related to each other by rotations about the six-fold symmetry axis of the hexagonal structure. Therefore, the two-dimensional structure of a manifold formed in such a way retains traces of positions' equivalence in the initial plane hexagonal crystal. The protein quasi_equivalence in the capsid achieved in this case significantly minimizes the number of conformations necessary for protein molecules for chemical bonding with their neighbors.

The geometrical peculiarities of the CK model lead to certain selection rules for the total number $N$ of proteins in the capsid: $N = 60T$, where $T = h^2 + k^2 + hk$, $h$ and $k$ are nonnegative integers. The protein arrangement order in the capsid, corresponding to the model, is explicitly used in deriving various biological and physical models [7–10]. At the same time, there is a number of exceptional virus families in which observed local order of protein arrangement is pentagonal, rather than hexagonal one [11]. An ordinary CK capsid contains 12 pentamers. Other $10(T - 1)$ capsomeres are hexamers. However, capsids of the papovavirus family consist of 72 pentamers located, according to common opinion, at all nodes of the CK spherical lattice with $T = 7$. According to the CK theory, such a capsid should contain 12 pentamers and 60 hexamers, i.e., 420 protein molecules, rather than 360 observed in fact experimentally. This is one of the known examples demonstrating the nonuniversality of the CK geometrical model as a model which makes proteins quasi-equivalent in the capsid. The quasi-equivalence can also be achieved based on completely different geometrical principles. For example, a structural model of papovavirus family capsids, based on the approach of spherical quasicrystalline tilings was proposed in [12, 13]. In this model, the spherical capsid surface was tessellated by two types of structural elements, i.e., tiles called the "kites" and "darts." Aperiodic tessellation of surface by several tile types is widely used to explain the quasicrystal structures [14]; the mathematical idea of this type of tiling was proposed by Penrose [15]. In the model [12, 13], proteins were put in tile corners with identical angles, the procedure being considered as a generalization of the CK quasi-equivalence principle.

However, despite the fact that the idea of the tessellation of spherical viral capsids of the papovavirus family into identical structural elements was first advanced in [12, 13], this idea remained almost undeveloped. In the Twarock model [12, 13], proteins were randomly arranged on the tile surface, rather than at nodes of the tiling. Furthermore, the proposed tiles were only approximately identical, which can be proved by the simplest geometrical analysis. All these facts worsened the protein quasi-equivalence in the model of [12, 13]. As shown below, the better protein quasi-equivalence can be achieved within the approach of spherical quasicrystalline tilings based on completely different geometrical principles.

The aim of the present work is the development of the approach based on the quasi-equivalence principle according to which the sphere tiling with the minimal number of different tile types, also minimizes the number of conformations necessary for identical protein molecules for bonding with their neighbors. In addition, we assume that protein molecules are located at vertices of tiles with identical edges, and we show how capsid structures of several well-known viruses can be interpreted within this approach. These viruses are Satellite Tobacco Mosaic Virus, $L–A$ Virus, Dengue Virus, Chlorosome Vigna Virus, Sindbis Virus, and Bovine Papilloma Virus of the papovavirus family.

## 2. BASIC PRINCIPLES OF THE PROPOSED APPROACH

In this work we develop the previously proposed model [16] of capsids of the papovavirus family, based on the dodecahedral, rather than icosahedral, capsid net. In [16], each pentagonal capsid face was decorated by the pentagonal Penrose quasilattice reconstructed in a chiral way. In addition, proteins were arranged at vertices of tiling consisting of three tile types, i.e., pentagons, narrow and wide rhombs. The minimum set of different tiles presented in the structure obviously makes it compatible with the quasi-equivalence principle; its practical implementation in nature is promoted by that fact that within the CK geometrical model it is simply impossible to organize packing of 360 proteins. In addition to [16], the geometrical ideas proposed in the present paper allow us to write the equations necessary to calculate the coordinates of positions occupied by proteins in capsids.

Note that, in its general form, the mathematical problem of the sphere surface tessellation by several types of identical structural elements is extremely complex. However, there are two facts which simplify greatly the construction of the simplest tessellations containing a limited number of tiles. First, we consider only symmetric spherical tilings with the icosahedral symmetry $I$, corresponding to all rotations of the icosahedron symmetry group. Second, the approximate positions of vertices of the structures we are looking for are in fact known: these are positions of protein molecule centers of mass. Because of the molecule asymmetry, proteins cannot find themselves on symmetric directions of the group I; therefore, each packing can be characterized by a certain integer $T$ equal to the number of nonequivalent by symmetry 60-fold general positions in the structure. Consequently, the tessellation always contains $60T$ positions.

Since the positions of considered tilings are situated on the sphere, the structure with a fixed $T$ is determined by $2T$ algebraic equations only. As a first approximation of the solutions of these equations, it is convenient to take approximately known position of centers of mass of protein molecules. If it is possible to solve these equations, and the coordinates of positions obtained differ only slightly from the initial ones, then the target tessellation exists.

Below, the first six simplest tilings ($T$ = 1, 2, 3, 3, 4, 6) with icosahedral symmetry I, corresponding to the well-known viral capsid structures are constructed (see Fig. 1, left column). The first, fourth, and fifth capsids satisfy the CK geometrical model; the second, third, and sixth capsids do not satisfy it, since their structural organization is completely different. The central column of Fig. 1 shows the approximately determined positions of centers of mass of protein molecules [16, 17]. Note that to determine these positions for the capsid shown in the sixth line, we used the positions of maxima of the experimentally determined distribution density function of proteins in this Capsid [11]. The right column of Fig. 1 shows the idealized tilings; peculiar features of their construction are described in more detail in what follows; there we also briefly characterize the equations which should be formulated for determining the idealized coordinates of proteins in these model capsids.

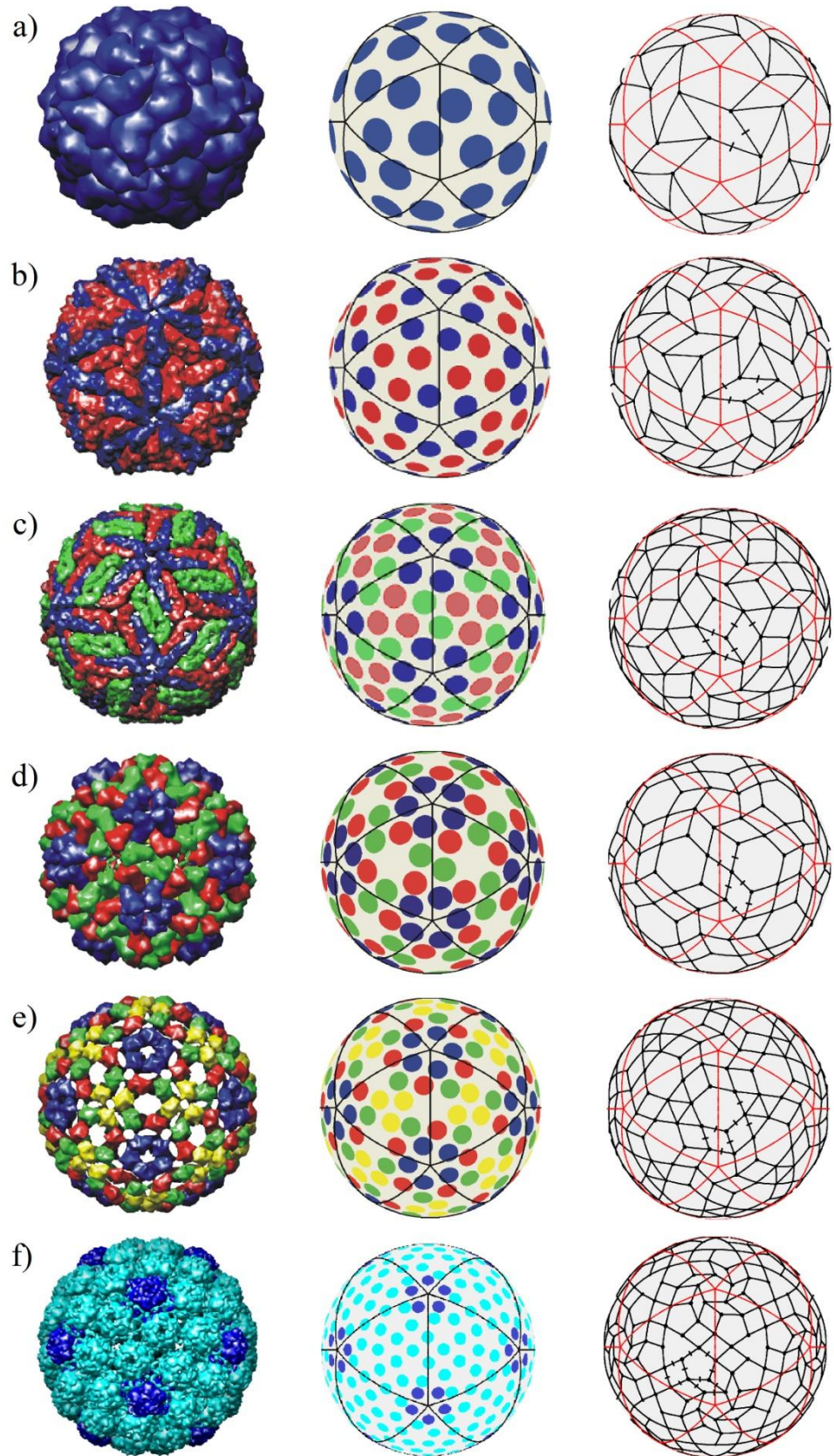

**Fig. 1.** Experimental structures of small viral capsids [18], positions of protein molecule centers of mass in them [16, 17], and regular spherical tilings with icosahedral symmetry $I$: (a) Satellite Tobacco Mosaic Virus $T = 1$; (b) $L–A$ Virus $T = 2$; (c) Dengue Virus $T = 3$; (d) Chlorosome Vigna Virus $T = 3$; (e) Sindbis Virus $T = 4$; and (f) Bovine Papilloma Virus $T = 6$. One set of symmetrically nonequivalent edges is selected in each figure. Edges of spherical icosahedra are indicated in the structures shown in middle and right columns. The minimum sets of symmetrically nonequivalent edges which should become equal in idealized tessellations are indicated by short bars.

## 3. CONSTRUCTION OF IDEALIZED SPHERICAL TILINGS

Let us note that the knowledge of symmetry features of the structures under consideration simplifies greatly the construction of spherical tilings. For example, for the capsid consisting of 60 protein molecules, one should know initial coordinates of only one position. Coordinates of other positions are determined from the initial ones using the vector representation matrices of the group I. To obtain matrices of all elements of this group, it is sufficient to add to 12 elements of the tetrahedron rotation group $T$ one more generator, e.g., rotation about the five-fold axis C5 shown in Table 1. Matrices of the group $T$ should be taken in the cube setting, where the two-fold axes of the group $T$ are parallel to coordinate system axes.

**Table 1.** Explicit form of the group I rotation matrices used in this study (Matrix $\mathbf{C_2}$ corresponds to rotation about the $z$ axis by an angle of $\pi$. $\mathbf{C_3}$ and $\mathbf{C_3'}$ are the matrices of rotations by an angle of $2\pi/3$ about the three_fold axes, $(1, 0, \tau^2)$ and $(-1, 0, \tau^2)$, respectively. Matrix $\mathbf{C_5}$ describes rotation about the five-fold axis $(0, 1, \tau)$ by an angle of $2\pi/5$. $\tau$ is the golden mean defined as $\tau=2\cos(\pi/5)=(\sqrt{5}+1)/2$.

| $\mathbf{C_2}$ | $\mathbf{C_3}$ | $\mathbf{C_3'}$ | $\mathbf{C_5}$ |
|---|---|---|---|
| $\begin{pmatrix} -1 & 0 & 0 \\ 0 & -1 & 0 \\ 0 & 0 & 1 \end{pmatrix}$ | $\dfrac{1}{2}\begin{pmatrix} 1-\tau & -\tau & 1 \\ \tau & -1 & 1-\tau \\ 1 & \tau-1 & \tau \end{pmatrix}$ | $\dfrac{1}{2}\begin{pmatrix} 1-\tau & -\tau & -1 \\ \tau & -1 & \tau-1 \\ -1 & 1-\tau & \tau \end{pmatrix}$ | $\dfrac{1}{2}\begin{pmatrix} \tau-1 & -\tau & 1 \\ \tau & 1 & \tau-1 \\ -1 & \tau-1 & \tau \end{pmatrix}$ |

To determine the coordinates of positions of the first idealized tiling shown in Fig. 1a, we should solve the following system of equations

$$\begin{cases} |\mathbf{C_5 r_1 - r_1}| = |\mathbf{C_2 r_1 - C_5 r_1}|, \\ |\mathbf{C_5 r_1 - r_1}| * \tau = |\mathbf{C_2 r_1 - r_1}| \end{cases} \quad (1)$$

where $\mathbf{r_1} = (x_1, y_1, z_1)$ are the coordinates of the target symmetrically nonequivalent position which determines the structure under consideration. The first equation expresses the equality of symmetrically nonequivalent sides of a narrow rhomb; the second equation sets the angle between them and follows directly from the law of cosines (see Fig. 2a). The tessellation in which this angle is equal to the typical "quasicrystalline" value $\pi/5$ is the closest to the experimental capsid structure.

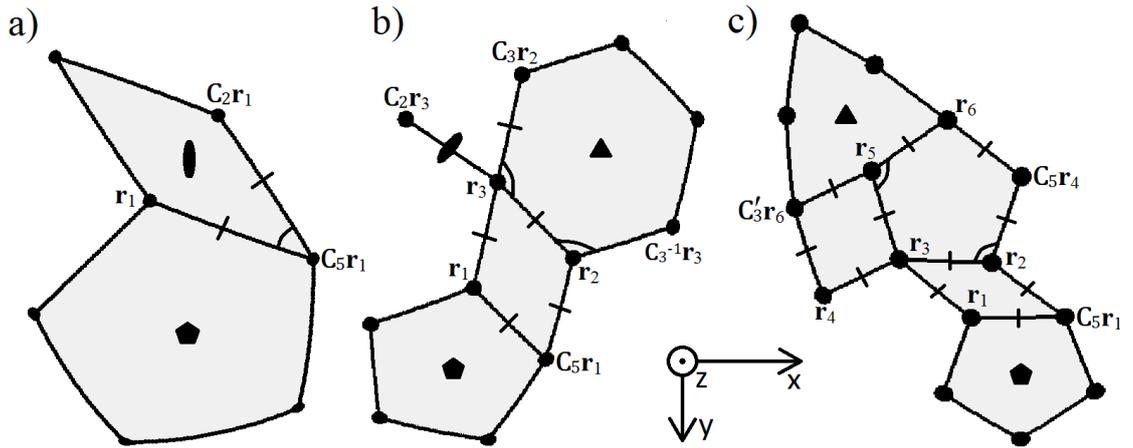

**Fig. 2.** Fragments of spherical tilings shown in Figs. 1a, 1d, and 1f, with the notations of symmetrically nonequivalent edges and angles, used in deriving the equations in systems (1)–(3).

Other idealized tessellations can be obtained in a similar way; only the number of equations describing the equality of symmetrically nonequivalent edges increases (see Figs. 1 and 2) in this case. For example, tiling with $T = 2$ is set by four equations expressing the equality of all four symmetrically nonequivalent sides of the narrow rhomb and the requirement that its acute angle would also be equal to $\pi/5$. The structure of the third (not Caspar–Klug) capsid with $T = 3$ is defined by six equations, five of which follow from the equality of six symmetrically nonequivalent edges of the corresponding tiling, and the sixth equation sets the acute angle of the narrow rhomb to be equal to $\pi/5$.

To determine the coordinates of three symmetrically nonequivalent positions of the fourth (Caspar–Klug) tiling, the following system of equations should be solved

$$\begin{cases} |\mathbf{C}_5\mathbf{r}_1 - \mathbf{r}_1| = |\mathbf{r}_3 - \mathbf{r}_1|, \\ |\mathbf{C}_5\mathbf{r}_1 - \mathbf{r}_2| = |\mathbf{r}_3 - \mathbf{r}_1|, \\ |\mathbf{r}_2 - \mathbf{r}_3| = |\mathbf{r}_3 - \mathbf{r}_1|, \\ |\mathbf{C}_2\mathbf{r}_3 - \mathbf{r}_3| = |\mathbf{r}_3 - \mathbf{r}_1|, \\ |\mathbf{C}_3\mathbf{r}_2 - \mathbf{r}_3| = |\mathbf{r}_3 - \mathbf{r}_1|, \\ |\mathbf{C}_3^{-1}\mathbf{r}_3 - \mathbf{r}_3| = |\mathbf{C}_3\mathbf{r}_2 - \mathbf{r}_2|, \end{cases} \tag{2}$$

where $\mathbf{r}_n = (x_n, y_n, z_n)$, n=1..3 are the coordinates of the target symmetrically nonequivalent positions, defining the idealized tessellation, matrix $\mathbf{C}_3^{-1} = \mathbf{C}_3^T$ defines rotation by an angle of $-2\pi/3$ about the three-fold axis $(1, 0, \tau^2)$. The first five equations correspond to the equality of six symmetrically nonequivalent edges of tiling structural elements, the sixth equation corresponds to the requirement that two symmetrically nonequivalent internal angles of the spherical hexagon would be equal (Fig. 2b). Under these conditions, the hexagon centered at the three-fold axis becomes regular, as it is assumed in the CK model.

The fifth (Caspar–Klug) idealized tessellation with $T = 4$ is characterized by eight equations, seven equations expressing the equality of lengths of eight symmetrically nonequivalent edges of the tiling. However, it is impossible to make regular the hexagon with a center on the two-fold axis, since the number of independent degrees of freedom in this case is insufficient. The last equation that we use for construction of this tiling is the equation similar to the last equation in system (2), which is set by the condition of equality of two internal spherical angles in the hexagon.

In the frame of the proposed model, it is also possible to describe the capsid structure of the bovine papilloma virus [11]. The set of equations describing molecule positions is divided into two groups. The first group consisting of 10 equations is defined by mutual equality of 11 symmetrically nonequivalent tiling edges. The second group of equations corresponds to the requirement of equality of internal angles of the pentamer not lying on the global five-fold axis (Fig. 2c),

$$\begin{cases} |\mathbf{r}_6 - \mathbf{r}_5| * \tau = |\mathbf{r}_6 - \mathbf{r}_3|, \\ |\mathbf{r}_3 - \mathbf{r}_2| * \tau = |\mathbf{C}_5\mathbf{r}_4 - \mathbf{r}_2|, \end{cases} \tag{3}$$

where $\mathbf{r}_n = (x_n, y_n, z_n)$, n=1..6 are the coordinates of the target symmetrically nonequivalent positions which determine the tiling.

The coordinates of symmetrically nonequivalent positions, determined by solving the above equations under the condition that they are situated on the unit sphere are listed in Table 2.

**Table 2.** Coordinates of symmetrically nonequivalent positions, defining tilings shown in Fig. 1. Other (equivalent) positions are obtained by the symmetry group I operations

| $T$ | Virus name | Coordinates | |
|---|---|---|---|
| 1 | Satellite Tobacco Mosaic Virus | (-0.09600, 0.12173, 0.98791) | |
| 2 | *L–A* Virus | (-0.21466, 0.30963, 0.92631); | (-0.17344, 0.09491, 0.98026) |
| 3 | Dengue Virus | (-0.09286, 0.70188, 0.70622); | (-0.26083, 0.71339, 0.65042); |
| | | (-0.09840, 0.87369, 0.47642) | |
| 3 | Chlorosome Vigna Virus | (0.05161, 0.32707, 0.94359); | (0.28665, 0.25587, 0.92323); |
| | | (0.10844, 0.74550, 0.99130) | |
| 4 | Sindbis Virus | (-0.04021, 0.67651, 0.73533); | (-0.25864, 0.69086, 0.67514); |
| | | (-0.10288, 0.81656, 0.56801); | (0.07839, 0.88863, 0.45186) |
| 6 | Bovine Papilloma Virus | (-0.09688, 0.40917, 0.90730); | (-0.05541, 0.29686, 0.95331); |
| | | (-0.24285, 0.28963, 0.92582); | (-0.39654, 0.36352, 0.84297); |
| | | (-0.29844, 0.10975, 0.94810); | (-0.14535, 0,00581, 0,98936) |

### 4. CONCLUSIONS

Finally, we note that the analysis of small and medium viral capsids, performed in terms of spherical tilings, made it possible to determine general features of these spherical structures. The obtained spherical tilings are mostly formed by regular pentagons, narrow and wide rhombs with angles typical of plane dodecagonal and pentagonal Penrose tiling [15]. Differently shaped tiles perturbing the motif are always most distant from the global five-fold axes; near the axes local tilings are of the same type in all six above cases, with a pentagon surrounded by a chiral arrangement of rhombs at the origin. This motif for the most complex structure with $T = 6$ extends due to five additional pentagons whose centers belong to general positions. Note that the latter spherical structure was obtained using the dodecahedron net decorated by the chirally reconstructed plane pentagonal Penrose tiling [16]; the tiling self_assembly theory was considered in [19] in the frame of the Landau crystallization theory.

Simpler capsid structures can be considered in much the same way [19]. The present paper develops our previous works [16, 19] and proposes a group-theory method for determining coordinates of nodes in idealized spherical tilings corresponding to structures of small and medium spherical viral capsids with icosahedral symmetry.

The closeness of the obtained idealized tilings to the structures of considered capsids shows that the proposed approach is general for many structures of small and medium viruses with the number of proteins $N \leq 360$. In contrast to capsids shown in lines (a), (d), and (e) of Fig. 1, the structures of capsids of the Papovavirus family ($T = 6$), $L$–$A$ Virus ($T = 2$), and Dengue Virus ($T = 3$) cannot be obtained within the geometrical scheme developed by Caspar and Klug. Therefore, the proposed approach is equally applicable to describe viral capsids both satisfying and not satisfying CK theory. At the same time, we do not reject the CK principle of capsid protein quasi-equivalence, but only propose the other geometrical method for implementing this principle. We emphasize that, in contrast to the CK theory where pentamers are topological defects formed during capsid assembly from a plane net, pentamers in the developed approach are intrinsic regular elements of spherical tilings. The minimum number of different tile types in the capsid provides quasi_equivalence of proteins and simplifies the internal structure of protein molecules, which, in turn, makes it possible to simplify the viralgenome and make it more compact. The presence of pentagonal tiles in the spherical tiling with global icosahedral structure is necessary. Such tiles are always located on the five_fold axes. However, the centers of much the same tiles can also form a 60-fold general position of the group I. Thus, capsids of the Papovavirus family containing 72 pentamers (instead of 12 as required by the CK model geometry) appear not an exotic anomaly, but the regular last term of the series of capsid structures with smaller numbers of proteins.

### ACKNOWLEDGMENTS

V.L. Lorman acknowledges the support of the Labex NUMEV.
This study was supported by the Russian Foundation for Basic Research (project no. 13_02_12085ofi_m).